\documentclass[aps,nofootinbib,prd,eqsecnum,twocolumn,showpacs,showkeys,preprintnumbers]{revtex4-1}

\usepackage{textcomp}
\usepackage{amsmath}
\usepackage{amssymb}
\usepackage{esint}

\makeatletter

\usepackage{graphicx}
\usepackage{amsfonts}
\usepackage{color}
\usepackage{bm}
\usepackage{mathrsfs}
\usepackage{epstopdf}
\usepackage{url}
\usepackage{footnote}
\usepackage{dsfont}
\usepackage{ulem}

\newcommand*{\rom}[1]{\expandafter\@slowromancap\romannumeral #1@}

\newcommand{\be}{\begin{equation}}
  \newcommand{\ee}{\end{equation}}
\newcommand{\ben}{\begin{eqnarray*}}
  \newcommand{\een}{\end{eqnarray*}}
\newcommand{\bea}{\begin{eqnarray}}
  \newcommand{\eea}{\end{eqnarray}}
\newcommand{\bdm}{\begin{displaymath}}
  \newcommand{\edm}{\end{displaymath}}
\newcommand{\ba}{\begin{align}}
  \newcommand{\ea}{\end{align}}

\makeatother

\begin{document}

\title{Electrodynamics and spacetime geometry: Astrophysical applications}

\author{Francisco Cabral}
\email{fc30808@alunos.fc.ul.pt}
\affiliation{Instituto de Astrof\'{\i}sica e Ci\^{e}ncias do Espa\c{c}o, Faculdade de
Ci\^encias da Universidade de Lisboa, Edif\'{\i}cio C8, Campo Grande,
P-1749-016 Lisbon, Portugal.}

\author{Francisco S. N. Lobo}
\email{fslobo@fc.ul.pt}
\affiliation{Instituto de Astrof\'{\i}sica e Ci\^{e}ncias do Espa\c{c}o, Faculdade de
Ci\^encias da Universidade de Lisboa, Edif\'{\i}cio C8, Campo Grande,
P-1749-016 Lisbon, Portugal.}

\date{\today}

\begin{abstract}
After a brief review of the foundations of (pre-metric) electromagnetism, we explore some physical consequences of electrodynamics in curved spacetime. 
  In general, new electromagnetic couplings and related phenomena are induced by the spacetime curvature. The applications of astrophysical interest considered here correspond essentially to the following geometries: the Schwarzschild spacetime and the spacetime around a rotating spherical mass in the weak field and slow rotation regime. In the latter, we use the Parameterised Post-Newtonian (PPN) formalism. 
We also explore the hypothesis that the   electric and magnetic properties of vacuum reflect the spacetime isometries. Therefore, the permittivity and permeability tensors should not be considered homogeneous and isotropic a priori.  
For spherical geometries we consider the effect of relaxing the homogeneity assumption in the constitutive relations between the fields and excitations. This affects the generalized Gauss and Maxwell-Amp\`{e}re laws where the electric permittivity and magnetic permeability in vacuum depend on the radial coordinate in accordance with the local isometries of space. For the axially symmetric geometries we relax both the assumptions of homogeneity and isotropy. 
We explore simple solutions and discuss the physical implications related to different phenomena such as: the decay of electromagnetic fields in the presence of gravity, magnetic terms in Gauss law due to the gravitomagnetism of the spacetime around rotating objects, a frame-dragging effect on electric fields and the possibility of a spatial (radial) variability of the velocity of light in vacuum around spherical astrophysical objects for strong gravitational fields.
\end{abstract}

\keywords{electrodynamics, (pseudo)-Riemann geometry, astrophysical applications}

\pacs{04.20.Cv, 04.30.-w, 04.40.-b}

\maketitle




\section{Introduction}

From the known physical interactions in Nature, the electromagnetic and gravitational forces were the first to be studied. They still continue to drive an incredibly rich theoretical and experimental research programs. These classical fields are the ones pervading our physical experience more directly at macroscopic scales, although the electromagnetic field is extremely relevant also at molecular, atomic and subatomic scales. On the other hand, gravity substantially governs the astrophysical and cosmological dynamics, but even at these scales electromagnetic fields are extremely important, in particular, in many astrophysical bodies and environments.
 Given the strong analogies between these interactions, it is not surprising that many historical attempts towards a unified field theory have occurred, driving important developments in theoretical physics \cite{unifiedfield}. Such a quest delves directly into one of the most fundamental questions in physics, namely the nature of space and time and its profound relation with physical fields and properties (see also \cite{Blagojevic:2012bc}).
 
It is well known that the theory describing electromagnetism can be expressed in the most general way using the exterior calculus of forms developed by Cartan. Based on previous works (see \cite{HehlYuribook} and references therein), Hehl and Obukhov  \cite{HehlYuribook,Gronwald:2005tv,Hehl:2000pe,Hehl:2005hu,Hehl:1999bt} clarified that this formalism is well established on four main postulates or axioms. These are charge conservation, magnetic flux conservation, Lorentz force and the constitutive relations between the fields ($\bold{E},\bold{B}$) and excitations ($\bold{D},\bold{H}$). The first three have a good experimental basis while the last postulate, which usually considers linear, local, homogeneous and isotropic relations, is not so well supported by experiment. These relations in vacuum can be understood as expressing constitutive relations for the spacetime itself \cite{Hehl:2005hu,Hehl:1999bt}. There is, one could say, a greater freedom in changing this postulate than the other three. Different electromagnetic models arriving from such changes are those of Heisenberg-Euler non-linear electrodynamics  \cite{HeisenbergEuler} or Mashhoon-Volterra non-local theory \cite{Mashhoon}, for example. In a previous work \cite{FCFL}, we briefly mentioned the idea of letting go the requirement of homogeneity and isotropy, which can be understood as a consequence of assuming that the electromagnetic properties of spacetime (or electrovacuum) have symmetries which follow the spacetime isometries (locally). 

Indeed, in the spirit of general relativity (GR), such isometries are not {\it a priori} given, instead they reflect the local geometry of specific physical situations. Likewise, the electric permittivity and magnetic permeability tensors should not be considered homogeneous and isotropic ({\it a priori}), regardless of spacetime local symmetries.  
Such alternative or extended electromagnetic theories can and should be tested experimentally in the context of the coupling between electromagnetic fields and gravity. Indeed, already at the foundational level there is an intimate link between spacetime geometry, in particular, the conformal structure, and electrodynamics via the constitutive relations \cite{Hehl:2005hu,Hehl:1999bt}. In this work, we start by considering linear, local, homogeneous and isotropic constitutive relations and study the effect of spacetime geometry on electromagnetic fields for specific physical situations. With this restriction on the constitutive relations we are then restricting the study to Maxwell electrodynamics in curved spacetime around astrophysical objects. We will also briefly study the effects of relaxing the condition of homogeneity for spherical (isotropic) geometries and relaxing both homogeneity and isotropy for the axial symmetric case.
The inhomogeneous equations, the Coulomb-Gauss and Oersted-Amp\`{e}re laws and the wave equations, constitute the basis for the physical applications we explore in the following sections. The equations will be shown explicitly and some physical consequences explored for two different geometrical backgrounds: 1) the Schwarzschild metric and 2) the geometry outside a stationary spherical gravitational mass with slow rotation in the weak field limit. (We considered the background of gravitational waves in a separate work \cite{FCFLGW}). The second case will be analysed using the Parameterized Post-Newtonian (PPN) formalism, first developed for solar system tests, which allows to parameterize different theories of gravity (see \cite{Efstathiou,Hartle,Landau,Schutz, Will:1971zza,Will:1972zz,Thorne:1970ws,Ni:1971wq,Misner:1974qy,Will:1981cz,Will:2005va}). 

As previously mentioned, both gravitational and electromagnetic fields are fundamental  at astrophysical scales governing the dynamics and driving the complex processes of structure formation (stars, galaxies, clusters, etc) as well as creating the conditions for extreme thermodynamical states of matter and nuclear reactions. Many highly energetic phenomena in astrophysics involve very strong gravity and electromagnetic fields interacting with very hot relativistic plasmas (see \cite{Brodin:2000du}). Therefore, the study of the coupling between gravity and electromagnetic fields is fundamental to a deeper understanding of many observed phenomena in high energy astrophysics. It is also relevant for the processes behind the formation and evolution of Active Galactic Nuclei (AGN) and to the physics of compact objects such as pulsars and black holes. For example, it is well know that gravitomagnetic fields play an important role in models for the collimation of astrophysical jets along a well established axis \cite{Jets,gravitomagneticBH}, but such models should include the coupling of electromagnetic fields with gravity. Such couplings, in particular, with the gravitomagnetic term, might be useful to deepen the understanding of astrophysical jets in radio (active) galaxies.
Magnetic fields are particularly relevant, since they pervade the physical Cosmos at planetary, stellar, galactic and extragalactic scales with different magnitudes (see \cite{Barausse:2014tra} and references therein). These fields also play a vital role in the complex interaction between protostars and the environment of the hosting molecular giant clouds, which drives the processes of star formation and  constitute a fundamental ingredient for the understanding of (phenomenological) stellar formation rates in galaxies \cite{astrophysics}. 

The applications in the following sections can have some astrophysical relevance but the main purpose here is to illustrate the effect of spacetime geometry in electric and magnetic fields motivated by relatively realistic astrophysical scenarios. An appropriate treatment of magnetic fields in relativistic astrophysical situations (with strong gravity) would require the equations governing fluids and fields of magnetohydrodynamics in the background of an appropriate spacetime metric. Furthermore, for cases with very strong magnetic fields, which can contribute to the gravitational field, the coupled Einstein-Maxwell equations (or its generalizations) need also to be included in the analysis. Such procedures require appropriate numerical methods and simulations.

The structure of this work is as follows. In Section \ref{secII}, we briefly review the foundations of electrodynamics and its deep relation to spacetime geometry \cite{FCFL}, presenting the basic equations to be explored in the applications. In Section \ref{secIII}, we explore some physical implications for two important geometrical backgrounds, namely the Schwarzschild metric and the exterior geometry of a stationary spherical gravitational mass with slow rotation in the weak field limit.
In Section \ref{conclusion}, we discuss our results and conclude.

\section{General formalism}\label{secII}

\subsection{Pre-metric formalism plus spacetime constitutive relations}

Before considering Maxwell fields in curved spacetime let us recall the more general (premetric) formalism at the foundations of electrodynamics. In the pre-metric formalism of electrodynamics, charge conservation gives the inhomogeneous field equations while magnetic flux conservation is expressed in the homogeneous equations  \cite{HehlYuribook,Gronwald:2005tv,Hehl:2000pe,Hehl:2005hu,Hehl:1999bt}
\begin{eqnarray}
&&d\bold{J}=0 \quad  \Rightarrow \quad  d\bold{G}=\bold{J},
  \nonumber\\
&&d\bold{F}=0 \quad  \Rightarrow \quad  d\bold{A}=\bold{F}.\label{fieldeqsForms} 
\end{eqnarray}
These are fully general, coordinate free, covariant equations without the need of an affine or metric structure of spacetime. Here $\bold{J}$ is the charge current density 3-form, $\bold{G}$ is the 2-form representing the electromagnetic excitation, $\bold{F}$ is the Faraday 2-form and $\bold{A}$ is the electromagnetic potential 1-form. The operator $d$ stands for exterior derivative.  The equations above were presented in a symmetric way to emphasize that, technically, the field $\bold{G}$ is a potential for the charge $\bold{J}$ in complete analogy with $\bold{A}$, which is a potential for the field strength $\bold{F}$. 

What provides the metric structure are the constitutive relations between $\bold{G}$ and $\bold{F}$. We will assume linear, local, homogeneous and isotropic relations given by the expression
\begin{equation}
\bold{G}=\lambda_{0}\star\bold{F},\label{constitutiveForms}
\end{equation}
where dimensional analysis allows us to write $\lambda_{0}=\sqrt{\varepsilon_{0}/\mu_{0}}$ with $\varepsilon_{0}$ and $\mu_{0}$ being the electric permittivity and the magnetic permeability of vacuum, respectively. The Hodge star operator $\star$ introduces the metric and maps $k$-forms to $(n-k)$-forms, with $n$ representing the dimension of the spacetime manifold.
With this assumption, the inhomogeneous equations can then be written by
\begin{equation}
d(\star\bold{F})=\lambda_{0}^{-1}\bold{J},
\label{FMaxwelleqForms}
\end{equation}
or in terms of the potential by
\begin{equation}
d\star d\bold{A}=\lambda_{0}^{-1}\bold{J}.\label{EqpotentialForms}
\end{equation}
In component form the above constitutive relations (\ref{constitutiveForms}) are given by
\begin{equation}
{\cal G}_{\mu\nu}=\dfrac{\lambda_{0}}{2}\sqrt{-g}g^{\alpha\lambda}g^{\beta\gamma}\epsilon_{\mu\nu\lambda\gamma}F_{\alpha\beta},\label{constitutive4d}
\end{equation}
where $g^{\beta\gamma}$ is the inverse metric and $\epsilon_{\mu\nu\lambda\gamma}$ is the 4-dimensional Levi-Civitta pseudo-tensor. The factor $\sqrt{-g}g^{\alpha\lambda}g^{\beta\gamma}\epsilon_{\mu\nu\lambda\gamma}$ is conformally invariant which means that at the foundational level electromagnetic fields are coupled to the conformal part of the metric. As a result, the light cone can be derived from linear and local electrodynamics \cite{Hehl:2005hu}. The 2-form fields $\bold{G}$ and $\bold{F}$ can be separated into their electric and magnetic parts according to
\begin{equation}
\bold{F}=\bold{E}\wedge d\sigma+\bold{B}, \qquad \bold{G}=-\bold{H}\wedge d\sigma+\bold{D},\label{Maxfooliation}
\end{equation}
where $\bold{E}$ and $\bold{H}$ are the electric field and magnetic excitation 1-forms, while $\bold{B}$ and $\bold{D}$ are the magnetic field and electric excitation 2-forms, respectively. Having performed a spacetime foliation, these fields are defined on the 3-dimensional hypersurfaces orthogonal to a time direction. 

The field equations in (\ref{fieldeqsForms}) have a well established empirical foundation (charge and magnetic flux conservations) while the constitutive relations are essentialy a postulate. Hehl and Obukhov have studied in detail more general constitutive relations beyond the usual local and linear relations \cite{Hehl:1999bt}. In \cite{FCFL}, we suggested that the permittivity and permeability tensors for vacuum should reflect locally the spacetime isometries and should not be considered a priori homogeneous and isotropic. This hypothesis changes the constitutive relations and therefore the dynamical equations for the electromagnetic fields. In this work, we will also address these changes in specific physical applications.  With the constitutive relations the main axiomatic (foundational) formalism of electrodynamics is complete.

\subsection{Electrodynamics in curved spacetime}

We will consider a 4-dimensional spacetime manifold with pseudo-Riemannian geometry with a $(+---)$ signature. Greek letters are spacetime indices ranging from 0 to 3 while Latin indices are space indices from 1 to 3. To explore electromagnetism in this geometry we will use the tensor formalism in order to better identify specific effects due to the components of the metric, for example, gravitomagnetic terms arising from off-diagonal time-space components which are a characteristic of axially symmetric spacetimes.

Integration theory and geometrical methods in 3-dimensional space can be used to derive the following electromagnetic equations which emerge from the principles of charge conservation, magnetic flux conservation and Lorentz force \cite{HehlYuribook,Gronwald:2005tv,Hehl:2000pe,Hehl:2005hu,Hehl:1999bt}
\begin{eqnarray}
\partial_{j}D^{j}=\varrho, \qquad
\epsilon^{ijk}\partial_{j}H_{k}-\partial_{t}D^{i}=\jmath^{i}, 
	\nonumber  \\
\partial_{j}B^{j}=0,
\qquad
\epsilon^{ijk}\partial_{j}E_{k}+\partial_{t}B^{i}=0,\label{Maxwellsimple}
\end{eqnarray}
where $\varrho\equiv\sqrt{-g}\rho$ and $\jmath^{i}\equiv \sqrt{-g}j^{i}$. Here  $D^{j}$ and $B^{k}$ are the components of vector densities (related to $\bold{D}$ and $\bold{B}$ through the expressions $D^{k}=\dfrac{1}{2}\epsilon^{kij}D_{ij}$, $B^{k}=\dfrac{1}{2}\epsilon^{kij}B_{ij}$). These equations also follow from (\ref{fieldeqsForms}) using (\ref{Maxfooliation}). To solve these equations one needs to consider the constitutive relations. It is easy to see \cite{FCFL} that the linear  relations in (\ref{constitutive4d}) or (\ref{constitutiveForms}) imply that, in general, for the quantities in Eq. (\ref{Maxwellsimple}) we get the expressions $D=D\left[E,B\right]$ and $H=H\left[E,B\right]$, which include a mixing of electric and magnetic components and the influence of the spacetime metric. As a consequence, magnetic terms appear in the Gauss law for geometries with non-vanishing time-space metric components. This is easily seen by considering the more traditional 4-dimensional tensor formalism. We will consider this below. 

The constitutive relations in (\ref{constitutive4d}) are implicitly assumed in the usual electromagnetic action of the more traditional variational or gauge approach. The set of Maxwell's equations in curved spacetime (pseudo-Riemann geometry) are the well known expressions 
\begin{eqnarray}
\nabla_{\mu}F^{\mu\nu}&=&\partial_{\mu}F^{\mu\nu}+\dfrac{1}{\sqrt{-g}}\partial_{\mu}(\sqrt{-g})F^{\mu\nu}=\mu_{0}j^{\nu},
   \nonumber \\
\partial_{[\alpha}F_{\beta\gamma]}&=&0.\label{MaxeqRiem}
\end{eqnarray}
where $F^{\mu\nu}=g^{\alpha\mu}g^{\beta\nu}F_{\alpha\beta}$ and we have used in the inhomogeneous equations the general expression for the divergence of anti-symmetric tensors in pseudo-Riemann geometry, $\nabla_{\mu}\Theta^{\mu\nu}=\dfrac{1}{\sqrt{-g}}\partial_{\mu}\left( \sqrt{-g}\Theta^{\mu\nu}\right)$. 

With the definitions
\begin{equation}
F_{0k}\equiv \dfrac{E_{k}}{c},\qquad  F_{jk}\equiv -B_{jk}=-\epsilon_{ijk}B^{i}
,\label{magneticfield}
\end{equation}
the homogeneous equations are the usual Faraday and magnetic Gauss laws
\begin{equation}
\partial_{t}B^{i}+\epsilon^{ijk}\partial_{j}E_{k}=0,\qquad  \partial_{j}B^{j}=0,
\end{equation}
while the inhomogeneous equations can be separated into the generalized Coulomb-Gauss and Oersted-Amp\`{e}re laws. These are, respectively
\begin{eqnarray}
\alpha^{kj}\partial_{k}E_{j}+\gamma^{j}E_{j}-cg^{m\mu}g^{n0}\epsilon_{kmn}\partial_{\mu}B^{k}
    \nonumber  \\
-c\sigma^{mn0}\epsilon_{kmn}B^{k}
=\dfrac{\rho}{\varepsilon_{0}},
   \label{GaussGenerall}
\end{eqnarray}
and
\begin{eqnarray}
\dfrac{1}{c}\alpha^{\mu ji}\partial_{\mu}E_{j}+\dfrac{1}{c}E_{j}\xi^{ji} 
-\epsilon_{kmn}\partial_{\mu}B^{k}g^{m\mu}g^{ni}
  \nonumber  \\
-B^{k}\epsilon_{kmn}\sigma^{mni}=\mu_{0}j^{i}, \label{MaxAmpereGenerall}
\end{eqnarray}
with the following geometric coefficients
\begin{equation}
\alpha^{kj}\equiv\left(g^{0k}g^{j0}-g^{jk}g^{00}\right),\quad \alpha^{\mu ji}\equiv\left(g^{0\mu}g^{ji}-g^{j\mu}g^{0i}\right)  ,\nonumber 
\end{equation}
\begin{eqnarray}
\gamma^{j}&\equiv &\Big[\partial_{k}\left(g^{0k}g^{j0}-g^{jk}g^{00}\right)
    \nonumber  \\
&&+\dfrac{1}{\sqrt{-g}}\partial_{k}(\sqrt{-g})\left(g^{0k}g^{j0}-g^{jk}g^{00}\right)\Big], \nonumber
\end{eqnarray}
\begin{equation}
\sigma^{mn\beta}\equiv\left[\partial_{\mu}(g^{m\mu}g^{n\beta})+\dfrac{1}{\sqrt{-g}}\partial_{\mu}(\sqrt{-g})(g^{m\mu}g^{n\beta}) \right], \nonumber
\end{equation}
\begin{eqnarray}
\xi^{ji}&\equiv &\Big[\partial_{\mu}\left(g^{0\mu}g^{ji}-g^{j\mu}g^{0i}\right)
    \nonumber  \\
&&+\dfrac{1}{\sqrt{-g}}\partial_{\mu}(\sqrt{-g})\left(g^{0\mu}g^{ji}-g^{j\mu}g^{0i}\right)\Big]. \nonumber 
\end{eqnarray}

We will refer to these as simply the Gauss and Maxwell-Amp\`{e}re laws.
One sees clearly, that new electromagnetic phenomena is expected due to the presence of extra electromagnetic couplings induced by spacetime curvature. In particular, the magnetic terms in the Gauss law are only present for non-vanishing off-diagonal time-space components $g^{0j}$, which in linearized gravity correspond to the gravitomagnetic potentials. These terms are typical of axially symmetric geometries (see \cite{Rezzolla:2000dk,Bini:2000gh}).

For diagonal metrics, the inhomogeneous equations, i.e., the Gauss and 
Maxwell-Amp\`{e}re laws, can be recast into the familiar forms
\begin{equation}
\partial_{k}\tilde{E}^{k}=\dfrac{\varrho}{\varepsilon_{0}},\label{gausschangevar}
\end{equation}
\begin{equation}
\epsilon_{ijk}\partial_{j}\bar{B}^{iijjk}=\mu_{0}(\jmath^{i}+\jmath^{i}_{D}),\label{maxampdisplacement}
\end{equation}
where
\begin{equation}
\tilde{E}^{j}\equiv -g^{jj}g^{00}\sqrt{-g}E_{j}, \qquad \varrho\equiv \sqrt{-g}\rho,
\label{gausschangevar2}
\end{equation}
and
\begin{eqnarray}
\bar{B}^{iijjk}\equiv g^{ii}g^{jj}\sqrt{-g}B^{k},\qquad \jmath^{i}\equiv \sqrt{-g}j^{i},
  \nonumber \\
\jmath^{i}_{D}\equiv -\varepsilon_{0}\sqrt{-g}\left(g^{00}g^{ii}\partial_{t}E_{i}+c^{2}E_{i}\xi^{ii}\right). \label{maxdisplacement}
\end{eqnarray}
%
%
%
%
%
%
%
%
(These equations are special cases of the more general expression which can be obtained from (\ref{MaxeqRiem}), namely $\partial_{\mu}\tilde{F}^{\mu\nu}=\jmath^{\nu}$, where $\tilde{F}^{\mu\nu}\equiv\sqrt{-g}F^{\mu\nu}$ and $\jmath^{\nu}\equiv \sqrt{-g}j^{\nu}$.)

New electromagnetic effects induced by spacetime geometry include an inevitable spatial variability (non-uniformity) of electric fields whenever we have non-vanishing geometric functions $\gamma^{k}$, electromagnetic oscillations (therefore waves) induced by gravitational radiation and also additional electric contributions to Maxwell's displacement current in the generalized Maxwell-Amp\`{e}re law. Notice that the functions $\xi^{ii}$ vanish for stationary spacetimes but might have an important contribution for strongly varying gravitational waves (high frequencies), since they depend on the time derivatives of the metric. Moreover, besides these predictions, as previously said, for axially symmetric spacetimes gravitomagnetic effects induce magnetic contributions to the Gauss law, with even static magnetic fields as possible sources of electric fields.
%
These are physical, observable effects of spacetime geometry in electromagnetic fields expressed in terms of the extended Gauss and Maxwell-Amp\`{e}re laws which help the comparison with the usual inhomogeneous equations in Minkowski spacetime, making clearer the physical interpretations of such effects. 

Finally, we review the field equations in terms of the electromagnetic 4-potential which in vacuum are also useful for the issue of electromagnetic waves.
From (\ref{EqpotentialForms}) or (\ref{MaxeqRiem}), we get
\begin{equation}
\nabla_{\mu}\nabla^{\mu}A^{\nu}-g^{\lambda\nu}R_{\varepsilon\lambda}A^{\varepsilon}-\nabla^{\nu}\left( \nabla_{\mu}A^{\mu}\right)=\mu_{0}j^{\nu},  
\end{equation}
where $R_{\varepsilon\lambda}$ is the Ricci tensor. Using the expression for the (generalized) Laplacian in pseudo-Riemann manifolds, $\nabla_{\mu}\nabla^{\mu}\psi=
\dfrac{1}{\sqrt{-g}}\partial_{\mu}\left( \sqrt{-g}g^{\mu\lambda}\partial_{\lambda}\psi\right)$, and considering vacuum we arrive at
\begin{eqnarray}
\partial_{\mu}\partial^{\mu}A^{\nu}+
\dfrac{1}{\sqrt{-g}}\partial_{\mu}\left( \sqrt{-g}g^{\mu\lambda}\right)\partial_{\lambda}A^{\nu}
    \nonumber \\
-g^{\lambda\nu}R_{\varepsilon\lambda}A^{\varepsilon}-\nabla^{\nu}\left( \nabla_{\mu}A^{\mu}\right)=0,\label{Proca}   
\end{eqnarray}
which is a generalized Proca-like equation with variable (spacetime dependent) effective mass induced by the curved geometry. The second term in Eq. (\ref{Proca}) can also be written in terms of the Levi-Civita connection, through the formula $g^{\alpha\beta}\Gamma^{\lambda}_{\,\alpha\beta}=-\dfrac{1}{\sqrt{-g}}\partial_{\alpha}\left( \sqrt{-g}g^{\alpha\lambda}\right)$, valid in pseudo-Riemann geometry. In usual Proca-like wave equations there is no such a term dependent on the first derivative of the (massive) vector field. Similar terms appear for wave phenomena with longitudinal modes. For a diagonal metric in vacuum, we get
\begin{eqnarray}
\partial_{\mu}\partial^{\mu}A^{\nu}+
\dfrac{1}{\sqrt{-g}}\partial_{\mu}\left( \sqrt{-g}g^{\mu\mu}\right)\partial_{\mu}A^{\nu}
    \nonumber \\
-g^{\nu\nu}R_{\varepsilon\nu}A^{\varepsilon}-\nabla^{\nu}\left( \nabla_{\mu}A^{\mu}\right)=0,\label{wavepotential} 
\end{eqnarray}
with no contraction assumed in $\nu$. In general, and contrary to electromagnetism in Minkowski spacetime, the equations for the components of the electromagnetic 4-potential are coupled even in the (generalized) Lorenz gauge ($\nabla_{\mu}A^{\mu}=0$). On the other hand, for Ricci-flat spacetimes, the term containing the Ricci tensor vanishes. Naturally, the vacuum solutions of GR are examples of such cases. New electromagnetic phenomena are expected to be measurable, for gravitational fields where the geometry dependent terms in Eq. (\ref{Proca}) are significant.

The inhomogeneous equations (Gauss and Maxwell-Amp\`{e}re laws) constitute the basis for the physical applications we explore in the following section. The equations will be shown explicitly and some physical consequences explored for different geometrical backgrounds: 1) Schwarzschild metric and 2) the geometry outside a stationary spherical gravitational mass with slow rotation in the weak field limit. The second case will be analysed using the PPN formalism (first developed for solar system tests) which allows to parameterize different theories of gravity. The case corresponding to the background of gravitational waves was presented in a separate work \cite{FCFLGW}.

\section{Applications}\label{secIII}

\subsection{Electromagnetic fields in spherically symmetric spacetime}
\label{secIIIA}

\subsubsection{The spherically symmetric spacetime geometry}

In this section, we consider the spherically symmetric spacetime around a spherical mass $M$, given by the Schwarzschild solution of Einstein's equations 
\begin{eqnarray}
ds^{2}=\left[1-\Psi(r)\right]c^{2}dt^{2}-\left[1-\Psi(r)\right]^{-1}dr^{2}
    \nonumber  \\
-r^{2}\left(d\theta^{2}+\sin\theta^{2}d\varphi^{2}\right),\label{Scw}
\end{eqnarray}
where $\Psi(r)\equiv 2GM/c^{2}r$.
This is the metric representing the geometry outside a (non-rotating) spherical mass due to a star for example, or outside a spherical (non-rotating) blackhole. When we consider the above metric in the next applications we assume that $r>2GM/c^{2}$. We recall that the coordinate $r$ in general has no direct relation to the physical (proper) distance. Rather, it was chosen such that given the line element $dl^{2}=f(t,r')d\Omega^{2}$ (where $d\Omega^{2}=d\theta^{2}+\sin\theta^{2}d\varphi^{2}$)
for the 2-spheres of constant $t$ and $r'$, under the appropriate coordinate change ${r'}\longrightarrow{r}$ (such that $f(t,r)=r^{2}$), the perimeter and area of the 2-spheres with constant $t$ and $r$, are given by the expressions $2\pi r$ and $\pi r^{2}$, respectively  \cite{Schutz}. 

Inside stars with radius $R_{*}$, imposing the continuity for the metric functions at $r=R_{*}$, we get
\begin{equation}
ds^{2}=e^{2\Phi(r)}c^{2}dt^{2}-\left[1-\psi(r)\right]^{-1}dr^{2}-r^{2}\left(d\theta^{2}+\sin\theta^{2}d\varphi^{2}\right),
\end{equation}
where $\psi(r)\equiv 2Gm(r)/c^{2}r$, $m(R_{*})\equiv M$ and the surface of the star is defined such that $p(R_{*})=0$, ($p$ is the pressure inside the star). The potential $\Phi(r)$ is obtained from the $r$ component of the energy-momentum conservation laws $\nabla_{\mu}T^{\mu\nu}=0$, which gives \cite{Hartle,Schutz}
\begin{equation}
(\rho_{m}c^{2}+p)\dfrac{d\Phi}{dr}=-\dfrac{dp}{dr},
\end{equation}
where $\rho_{m}$ is the mass density. To solve this equation one needs the equation of state describing the thermodynamical properties of the interior of the star, as well as the other two remaining differential equations describing the inner structure of spherical relativistic stars, which are derived from Einstein's equations. These correspond to the equations for $\rho_{m}(r)$ and $m(r)$.
The coordinate pathology happening for $r=2GM/c^{2}$ in the black hole case, has no similar problem here because a careful analysis of the interior solution for ordinary stars shows that $r>2Gm(r)/c^{2}$ always \cite{Schutz}. 
A simple pedagogical model which is not realistic (in fact it predicts an infinite speed of sound inside the star) is that of a star with constant density. Other more realistic well known exact solution is that of Buchdahl (1981) \cite{Schutz}, which assumes an equation of state of the form $\rho_{m}/c^2=12\sqrt{\bar{p}p}-5p$, where $\bar{p}$ is an arbitrary constant, which can be made causal (with a speed of sound lower than that of light) and which for low pressures reduces to the equation of state of a $n=1$ polytrope in Newtonian theory of stellar structure \cite{Schutz}.
 
Beyond ordinary stars, some old neutron stars can have a negligible rotation and have an approximately spherical gravitational metric field both inside and outside the matter. Nevertheless, the theoretical research on the equation of state for neutron stars is more delicate and it still has many open questions due to our lack of understanding of the properties of matter at supranuclear densities characteristic of the inner cores of such compact objects. We hope to clarify many of the physical issues involved with the advent of precision gravitational wave astronomy, in particular with its applications to asteroseismology. For simplicity, we will not consider interior solutions. 
 
We now proceed with the analysis of electrodynamics in the background of the spherical geometry in Eq. (\ref{Scw}). 
 Let us start by observing that in general, the first term in Maxwell's inhomogeneous equations in Eq. (\ref{MaxeqRiem})
for any fixed $\nu$, corresponds to the usual divergence, while the factors in the second term which are being contracted with (the contravariant components of) the Faraday tensor correspond to the components of a gradient. These terms must be computed using the appropriate expressions for a given system of coordinates. From these considerations in spherical coordinates, with the metric in Eq. (\ref{Scw}), the following expressions are obtained for the geometric functions that enter in the Gauss law (\ref{GaussGenerall})
\begin{equation}
\gamma^{r}(\bold{x})=\dfrac{4}{r},
\qquad 
\gamma^{\theta}(\bold{x})=\dfrac{2\cot \theta}{r^{3}\left[1-\Psi(r)\right]},
\qquad \gamma^{\varphi}(\bold{x})=0,
\end{equation}
and in the Maxwell-Amp\`{e}re law (\ref{MaxAmpereGenerall})
\begin{eqnarray}
\sigma^{r\varphi\varphi}(\bold{x})= \dfrac{1}{r^{3}\sin^{2}\theta}\left[2-\Psi(r)\right],\qquad
\sigma^{r\theta\theta}(\bold{x})=\dfrac{2-\Psi(r)}{r^{3}},
   \nonumber  \\
 \sigma^{\theta rr}(\bold{x})=\dfrac{2\cot \theta}{r^{3}}\left[1-\Psi(r)\right], \qquad
 \sigma^{\theta\varphi\varphi}(\bold{x})=0, \nonumber \\
%
\sigma^{\varphi\theta\theta}(\bold{x})=0,\qquad
\sigma^{\varphi rr}(\bold{x})=0,\qquad \xi^{ii}(\bold{x})=0,\nonumber
\end{eqnarray}
respectively.

\subsubsection{Electrostatics in the Schwarzschild geometry}
 
The Gauss law (\ref{GaussGenerall}) in the Schwarzschild geometry (\ref{Scw}), for the case of a static radial field with spherical symmetry becomes
\begin{equation}
\dfrac{dE_{r}}{dr}+\frac{4}{r}E_{r}=\dfrac{\rho}{\varepsilon_{0}}.
\end{equation}
The charge distribution can be that of a spherical charge or a spherical shell. In any case, a characteristic radius can be defined, which we simply represent by $R$. In vacuum, outside the gravitational source, be it a star or a black hole (in which case we consider $r>2GM/c^{2}$) and outside the charge distribution, the solution to Gauss' law is
\begin{equation}
E_{r}=\dfrac{C}{r^{4}}, \label{electricSch}
\end{equation}
contrasting with the simple $r^{-2}$ dependence in usual electrostatics. Through dimensional analysis we can write $C=\alpha kQR^{2}$,
%
%
where, as mentioned before, $R$ can be viewed as the radius of the charged body, $k\equiv1/4\pi\varepsilon_{0}$ and $\alpha$ is an arbitrary (dimensionless) constant which can be fixed by comparing with the inner solution at $r=R$ and imposing continuity.
Near the surface of the massive body the deviation of our solution with respect to the case without gravity is largest. In particular, such effects can be considered for charged stars or charged black holes, by identifying $R$ with $R_{*}$ in the case of stars or with $2GM/c^{2}$ for the black hole case.

One can also assume, as an interesting exercise, the case in which the charge distribution is due to a spherical shell (with outer radius $R$) immersed in the Schwarzschild geometry (\ref{Scw}). In such a case, inside the spherical charged body but still outside the black hole or outside the star, by assuming uniform charge density, the solution of Gauss' law is given by
\begin{equation}
E_{r}=\dfrac{\tilde{C}}{r^{4}}+\dfrac{\rho}{5\varepsilon_{0}}r,
\end{equation}
where $2GM/c^{2}<r<R$ for black holes, and $R_{*}<r<R$ for stars.
Recall that in usual electromagnetism inside spherical insulators we have  $E_{r}=r\rho/3\varepsilon_{0}$ plus a vacuum term (proportional to an arbitrary constant) which is usually chosen to be zero to guarantee continuity at $r=R$. Therefore, in usual electrostatics, both solutions coincide for $r=R$.  Similarly, setting the vacuum term to zero in the above solution ($\tilde{C}=0$) and imposing the continuity of the solutions at $r=R$, fixes the value of $\alpha$. We conclude that in the Schwarzschild geometry, the electric field decays more strongly outside a (spherical) charged body and increases more slowly inside, in comparison with the same physical situation in flat spacetime. In particular in vacuum, we get 
 \begin{equation}
\dfrac{E_{r}^{flat}(r)-E_{r}^{curved}(r)}{E_{r}^{flat}(r)}=1-\alpha\dfrac{R^{4}}{r^{4}},\qquad r\geq R.\label{electricsch2}
\end{equation}


\subparagraph{Electric field due to charged plates in equatorial orbit in the (gravitational) weak field limit:}

Previously, we assumed that the electric field had spherical symmetry just like the metric. Now we consider the simplest case in which the electric field has a single component $E_{x}$ (in a certain fixed system of coordinates). In this case, outside the charge distribution, the Gauss law (\ref{GaussGenerall}) provides
\begin{equation}
\partial_{x}E_{x}=\dfrac{\gamma^{x}}{g^{xx}g^{00}}E_{x}.
\label{electricnonunif}
\end{equation}
Let us consider the weak field limit of the Schwarzschild solution (expanding the metric in powers of $2GM/c^{2}r$ up to first order). In isotropic Cartesian coordinates the metric is then given by
\begin{eqnarray}
ds^{2}=c^{2}\left(1-\dfrac{2GM}{c^{2}\tilde{r}}\right)dt^{2}-\left(1+\dfrac{2GM}{c^{2}\tilde{r}}\right) \times
   \nonumber \\
\times\left( dx^{2}+dy^{2}+dz^{2}\right),
\end{eqnarray}
where, $\tilde{r}\equiv\sqrt{x^{2}+y^{2}+z^{2}}$.
We then have the following solution
\begin{equation}
E_{x}=E_{0x}e^{\int _{x_{0}}^{x}\gamma^{x}g_{xx}g_{00}dx}=E_{0x}\sqrt{\dfrac{\tilde{r}/R_{sch}-1}{\tilde{r}/R_{sch}+1}}  \,.
\label{Explanecap}
\end{equation}
In flat spacetime, the electric field is uniform with magnitude given by $E_{0x}$. Sufficiently far away from the gravitational source (star, planet or black hole) the electric field is uniform. The effect of gravity on the electric field can then be characterized by the dimensionless quantity
\begin{equation}
\vert E_{x}-E_{0x}\vert/E_{0x}=1-\sqrt{\dfrac{\tilde{r}/R_{sch}-1}{\tilde{r}/R_{sch}+1}},
\end{equation}
which vanishes when $\tilde{r}\rightarrow\infty$.
 Recall that electric fields with quasi parallel field lines can result from charged plates. In particular, this result means that in the vicinity of a spherical mass, the electric field created by a single charged plate or by a plane capacitor (oriented along the $x$ axis in this case) is no longer uniform. The field changes its magnitude due to spacetime curvature. Far from the gravitational source we recover the uniform electric field.

An interesting application of this result is the following thought experiment: Consider a plane capacitor with charge $Q$, an area given by $A=L^{2}$ and a distance between plates $D$. Suppose this system is put in equatorial orbit around a certain quasi-spherical astrophysical mass $M$ with negligible rotation. Further, suppose that the line perpendicular to the capacitor's plates is always aligned with a certain reference distant Quasar, in the $x$ axis (of the Cartesian system centred in the gravitational source). Once the system is in orbit, that direction will be aligned with the radial direction relative to the central mass, twice per cycle. Alternatively this direction will be perpendicular to the radial line, also twice a cycle (when the system is crossing the $y$ axis). The electric field, which is always aligned with the $x$ axis, has a magnitude which varies with the distance to the center and therefore, twice a cycle the change in magnitude is either along the line connecting the two plates or along the perpendicular to that direction. The maximum change in magnitude for both cases is equal if $D=L$. On the other hand, if $D \gg L$ and $L$ is sufficiently small, then the field is approximately uniform whenever the system is crossing the $y$ axis and non-uniform in the rest of the orbit. If, whenever the capacitor is crossing the $x$ axis the maximum change in magnitude is $\vert\Delta E_{x}\vert$, then the dimensionless quantity of experimental relevance is then
$\vert\Delta E_{x}\vert/E_{0x}$, which measures the strength of the effect of gravity in the weak field regime. The value $E_{0x}=Q/\varepsilon_{0}L^{2}$ is the usual value of the uniform field inside the capacitor in the absence of gravity.

The strength of this effect when the capacitor is crossing the $x$ axis, with one plate at position $x$ and the other at $x+D$, is given by
\begin{equation}
\dfrac{\vert E_{x}(x)- E_{x}(x+D)\vert}{E_{0x}}=\sqrt{1-\dfrac{2}{1+\tilde{X}+\tilde{D}}}-\sqrt{\dfrac{\tilde{X}-1}{\tilde{X}+1}},
\end{equation} 
where the distances are in units of $R_{sch}$, $\tilde{X}\equiv \tilde{r}/R_{sch}$, $\tilde{D}\equiv D/R_{sch}$. This dimensionless quantity measures the maximum change of the magnitude of the electric field inside the capacitor due to the gravitational field of the astrophysical spherical mass. In principle, this effect could be used to test GR and modified theories of gravity, providing another complementary (weak field) test in the Solar System.

Since the values of the Schwarzschild radius for the Sun and Jupiter are approximately $2.95\times 10^{3}$m and $2.2$m, respectively, the effect should be very small unless one gets extremely close to their surfaces. For example, for a laboratory in orbit around Jupiter at a distance approximately equal to three times Jupiter's radius ($\tilde{X}\sim 9\times 10^{7}$) we have the following values 
\begin{eqnarray}
\vert\Delta E_{x}\vert/E_{0x}\sim10^{-14}\quad (\tilde{D}=100)\,,
   \nonumber  \\
\vert\Delta E_{x}\vert/E_{0x}\sim10^{-13}\quad (\tilde{D}=1000).
\end{eqnarray} 

Naturally, these tiny values represent an enormous experimental challenge in terms of the sensitivities and noise control requirements. The fact that we are not using test masses, but electric fields instead to study gravity, complicates further the experiment due to various possible environmental effects related to space weather, in particular, solar and planetary magnetospheres. In any case, it is always better to measure voltage differences than to measure the electric field itself, since better sensitivities can be obtained. The Voltage drop between the two plates will be less than in the absence of gravity. In principle, since different points inside the capacitor will be at different radial coordinates which change with time, we also expect the generation of electromagnetic radiation with a frequency related to the orbital frequency of the spacecraft. This electromagnetic radiation should be linearly polarized and is completely induced by the effect of gravity in the electric field inside the capacitor and the orbital motion of the spacecraft.

\subparagraph{Electric field in the gravitational field of a massive spherical object --  Case of non-homogeneous, isotropic constitutive relations:}
In \cite{FCFL} we considered the possibility of changing the constitutive relations between the electromagnetic fields and excitations by relaxing the assumptions of homogeneity and isotropy. The idea behind this suggestion comes from the very deep relation between spacetime geometry and electrodynamics, already present at the foundations of electromagnetism, and so well explained by Hehl and Obukhov. But, it also comes from the notion that the physical properties of vacuum (or electrovacuum) should not be a priori given. This follows the spirit of GR which is a background independent theory and therefore the local geometry of spacetime is not a priori given, rather it has to be considered for each physical system as a solution of the dynamical equations. Likewise, we postulate that the electric permittivity and magnetic permeability tensors for vacuum should reflect the local symmetries of spacetime geometry.   

Let us consider spherical symmetry. Relaxing the assumption of homogeneity around spherical bodies, the linear, local, isotropic and non-homogeneous constitutive relation between the electric field and electric excitation is \cite{FCFL}
\begin{equation}
D^{i}=-g^{00}g^{ij}\sqrt{-g}\,\varepsilon_{0}(r)E_{j}.\label{newconstitutiverel}
\end{equation}
The Gauss law can be written as
\begin{equation}
\partial_{i}D^{i}=\rho \sqrt{-g}.
\end{equation}

We will use the following Ansatz
\begin{eqnarray}
\varepsilon_{0}(r)=\varepsilon_{0}\left(1+\bar{\gamma}\dfrac{2GM}{c^{2}r}\right).
   \label{permitivityschw}
\end{eqnarray}
which can be seen as the linear approximation of a Taylor expansion in powers of $2GM/(c^{2}r)$. Here $\bar{\gamma}$ is a dimensionless parameter. Then, considering the Schwarzschild geometry, and choosing $\bar{\gamma}=1$ we find the following solution in vacuum
\begin{equation}
E_{r}=\dfrac{C}{r^{4}}\dfrac{1}{(1+R_{sch}/r)}. 
\end{equation}
This result includes the effect of a radial dependence of the electric permittivity of vacuum for spherical gravitational fields. The strength of the effect is given by
\begin{equation}
\vert E^{\varepsilon_{0}(r)}_{r}-E_{r}^{\varepsilon_{0}} \vert/E_{r}^{\varepsilon_{0}}=\dfrac{R_{sch}/r}{1+R_{sch}/r}\label{nonhomogeneityeffect1},
\end{equation}
where $E_{r}^{\varepsilon_{0}}$ corresponds to the solution in (\ref{electricSch}).

According to the hypothesis we explore here in the curved geometry around a massive object, the electric permittivity of vacuum is not homogeneous, but has a radial dependence instead. For black holes, this effect influences electric fields more strongly near the horizon. The strength of the effect is maximum at the horizon where the magnitude of the electric field is $0.5$ times weaker than the case with a constant permittivity tensor. 

If we now repeat the analysis for the case of the electric field inside a plane capacitor in orbit around a spherical mass, we arrive at the result
\begin{equation}
E_{\tilde{r}}=E_{0x}\dfrac{\sqrt{1-R_{sch}^{2}/\tilde{r}^{2}}}{(1-R_{sch}/\tilde{r})^{2}}.
\end{equation}  
The strength of the effect can be quantified again by comparison with the case where the electric permittivity is constant, where we get
\begin{equation}
\vert E^{\varepsilon_{0}(\tilde{r})}_{\tilde{r}}-E_{\tilde{r}}^{\varepsilon_{0}} \vert/E_{\tilde{r}}^{\varepsilon_{0}}=1-\dfrac{\tilde{r}/R_{sch}}{1+\tilde{r}/R_{sch}}.\label{nonhomogeneityeffect2}
\end{equation}
In principle, this effect could  be tested using electric fields inside capacitors and these expressions can be generalized to include the PPN approach.

\subsubsection{Magnetostatics in the Schwarzschild geometry -- Astrophysical applications}

Suppose that in some reference frame we have a static magnetic field due to a (stationary) current and no electric field. Then, the generalized Maxwell-Amp\`{e}re law in Eq. (\ref{Maxwellsimple}) is given by the following equations
\begin{equation}
\dfrac{1}{r\sin \theta}\partial_{\theta}\left(\sin \theta H_{\varphi}\right)-\dfrac{1}{r\sin\theta}\partial_{\varphi}H_{\theta}=\sqrt{-g}j^{r},
\end{equation}
\begin{equation}
\dfrac{1}{r\sin \theta}\partial_{\varphi} H_{r}-\dfrac{1}{r}\partial_{r}(rH_{\varphi})=\sqrt{-g}j^{\theta},
\end{equation}
\begin{equation}
\dfrac{1}{r}\partial_{r}(rH_{\theta})-\dfrac{1}{r}\partial_{\theta}H_{r}=\sqrt{-g}j^{\varphi}.
\end{equation}
We can solve this for the homogeneous and isotropic constitutive relations \cite{FCFL}, that can be derived from Eq. (\ref{constitutive4d}):
\begin{equation}
H_{k}=\mu_{0}^{-1}\sqrt{-g}g_{jk}B^{j}\,.
\end{equation}

The following applications can have some astrophysical relevance although, as previously mentioned, the main purpose here is to illustrate the effect of spacetime geometry in magnetostatics motivated from minimally realistic astrophysical scenarios. An appropriate treatment of magnetic fields in relativistic astrophysical situations (with strong gravity) would require the equations governing fluids and fields of magnetohydrodynamics in the background of an appropriate spacetime metric. Furthermore, for cases with very strong magnetic fields, which can contribute to the gravitational field, the coupled Einstein-Maxwell (or its generalizations) also need to be included in the analysis.  


\subparagraph{Magnetic field created by a ring of circulating plasma around a spherical mass:}
Consider a spherical astrophysical mass, which could be a black hole, with a stationary current loop around it on the equatorial plane ($\theta=\pi/2$). This current distribution, which could be due to a disk of very hot plasma, would create an axially symmetric magnetic field of the general form
\begin{equation}
\bold{B}=B^{r}(r,\theta)\bold{e_{r}}+B^{\theta}(r,\theta)\bold{e_{\theta}}\,,
\end{equation}
which implies
\begin{equation}
 \bold{H}=H_{r}(r,\theta)\bold{e^{r}}+H_{\theta}(r,\theta)\bold{e^{\theta}}.
\end{equation}
Then, the relevant Maxwell equation is
\begin{equation}
\dfrac{1}{r}\partial_{r}(rH_{\theta})-\dfrac{1}{r}\partial_{\theta} H_{r}=\sqrt{-g}j^{\varphi},
\end{equation}
which can be solved outside the current distribution by setting the RHS to zero. We will consider this equation in the equatorial plane and outside the disk (or ring) of currents, i.e., for $r > R$ where $R$ is a mean representative of the radius of the circular current distribution. Setting $\theta = \pi/2$ the magnetic field will only have the $\theta$ component 
\begin{equation}
\bold{B}=\pm B^{\theta}(r,\pi/2)\bold{e_{\theta}},
\end{equation}
where the $\pm$ refers to the cases in which the circulating current is moving in the direction of positive or negative $\varphi$, respectively. We then get
$H_{\theta}(r,\pi/2)=f(\pi/2)/r$, which implies
\begin{equation}
B^{\theta}(r,\pi/2)=\dfrac{\mu_{0}}{\sqrt{-g}\; g_{\theta\theta}}H_{\theta}\propto \dfrac{1}{r^{5}}.\label{magneticfieldschw1}
\end{equation}
We conclude that the magnetic field (on the equatorial plane) due to the circular current distribution decays faster with the radial distance, in the curved spacetime of Schwarzschild geometry, than in the flat (Minkowski) case (although, strictly speaking the radial coordinate here does not correspond to a physical distance). 

We recall that at this stage we are neglecting rotation, therefore these calculations can be viewed as having an approximate validity around quasi-static spherical masses with an electric current due to a highly ionized gas in the orbiting accretion disk. Since the electromagnetic equations were considered and solved outside the current distribution creating the magnetic field and in the exterior (Schwarzschild) spacetime, another astrophysical scenario compatible with the approach taken here is that of a (quasi) spherical mass with negligible rotation with a magnetic field generated by electric currents in its interior, as long as the magnetic field energy density has a negligible effect in the gravitational field (no back reaction).  


\subparagraph{Magnetic field created by an astrophysical jet:}
Consider now another idealized astrophysical scenario in which the spherical body emits a stationary jet of charged particles vertically, defining an axis. We set this to be the $z$ axis. In this case, a magnetic field will arise with axial symmetry and along the $\varphi$ direction with symmetrical configurations above or below the $\theta=\pi/2$ plane
\begin{equation}
\bold{B}=\pm B^{\varphi}(r,\theta)\bold{e_{\varphi}}.
\end{equation}
The relevant Maxwell-Amp\`{e}re equations outside the current distribution are
\begin{equation}
\dfrac{1}{r\sin \theta}\partial_{\theta}\left(\sin \theta H_{\varphi}\right)=0,\qquad -\dfrac{1}{r}\partial_{r}\left( rH_{\varphi}\right)=0,
\end{equation}
and therefore $H_{\varphi}(r,\theta)\propto (r\sin \theta)^{-1}$, which implies
\begin{equation}
B^{\theta}(r,\theta)=\dfrac{\mu_{0}}{\sqrt{-g}\;g_{\varphi\varphi}}H_{\varphi}\propto \dfrac{1}{r^{5}\sin^{4} \theta}.\label{magneticfieldschw2}
\end{equation}

Here we considered a constant magnetic permeability tensor for vacuum, but following our hypothesis that the properties of this tensor should reflect the local spacetime isometries it should be interesting to compute the changes to these results if the magnetic permeability has a radial dependence. This is what we do next.



\subparagraph{Magnetic field around a spherical gravitational field, with non-homogeneous constitutive relations:} 
Let us consider the effect of relaxing the condition of homogeneity in the constitutive relations, assuming a radial dependence of the permeability tensor in vacuum, i.e.,
\begin{equation}
H_{k}=\mu_{0}^{-1}(r)\sqrt{-g}g_{jk}B^{j} \,.
\label{contitutivemagnonhomog}
\end{equation}
 We introduce the ansatz
\begin{equation}
\mu_{0}(r)=\mu_{0}\left(1+\bar{\bar{\gamma}}\dfrac{r_{Schw}}{r}\right),\label{permeabilityschw}
\end{equation}
and choose $\bar{\bar{\gamma}}=1$.
The above results are therefore generalized into
\begin{eqnarray}
 B^{\theta}(r,\pi/2)&\propto & \dfrac{\mu_{0}}{r^{5}}\left(1+\dfrac{r_{Schw}}{r}\right)\,,
      \nonumber \\
 B^{\theta}(r,\theta)&\propto & \dfrac{\mu_{0}}{r^{5}\sin^{4} \theta}\left(1+\dfrac{r_{Schw}}{r}\right),
 \nonumber
\end{eqnarray}
for the magnetic field due to a ring of current (in the equatorial plane) and to the astrophysical jet, respectively. The strength of this effect in comparison with the case with a homogeneous magnetic permeability tensor, is stronger near the horizon
\begin{equation}
\dfrac{\vert \delta \bold{B}\vert}{\bold{B}}=\dfrac{r_{Schw}}{r}.
\end{equation}

Contrary to what happened with the electric field case, the radial dependence of the permeability tensor enhances the magnetic field in comparison with the case of magnetic homogeneity of vacuum. In the limit, at the horizon of black holes, the magnetic field is stronger by a factor of 2. This fact comes from the very nature of the constitutive relations and the physical dimensions of the magnetic field. Indeed, the magnetic permeability appears in the denominator in (\ref{contitutivemagnonhomog}) while the electric permittivity appears in the numerator in (\ref{newconstitutiverel}). Like in the case of the electric field, sufficiently far way from the source, the effect becomes neglibible.



\subsection{Maxwell fields around slowly rotating objects in the weak field limit within the PPN formalism}\label{secIIIB}

The appropriate metric describing the local geometry outside a stationary rotating mass is the Kerr metric \cite{kerrspaceVisser:2007fj}. Expanding this geometry to first order in the angular momentum $J$, we get the geometry outside the source in the limit of slow rotation \citep{Hartle}
\begin{equation}
ds^{2}_{Kerr}\approx ds^{2}_{Schawrzschild}+\dfrac{4GJ}{c^{2}r}\sin ^{2}\theta d\varphi dt.
\end{equation}
Recall that the deformations of the object's spherical symmetry depend quadratically with angular momentum $J$, while the metric already changes at the linear level. Therefore, this geometry is a good approximation to that of a slowly rotating (quasi) spherical gravitational mass.

In the weak field limit (non-relativistic weak gravitational sources) the above metric is given by the following expressions in spherical and (isotropic)  Cartesian coordinates, respectively \citep{Hartle}
\begin{eqnarray}
ds^{2}=c^{2}\left(1-\dfrac{2GM}{c^{2}r}\right)dt^{2}-\left(1+\dfrac{2GM}{c^{2}r}\right)dr^{2}
   \nonumber  \\
-r^{2}\left(d\theta^{2}+\sin\theta^{2}d\varphi^{2}\right)+\dfrac{4GJ}{c^{2}r}\sin ^{2}\theta d\varphi dt,
\end{eqnarray}
\begin{eqnarray}
ds^{2}=c^{2}\left(1-\dfrac{2GM}{c^{2}\tilde{r}}\right)dt^{2}-\left(1+\dfrac{2GM}{c^{2}\tilde{r}}\right) \times
    \nonumber \\
\times \left( dx^{2}+dy^{2}+dz^{2}\right)+\dfrac{4GJ}{c^{2}\tilde{r}^{3}}dt(xdy-ydx),
\end{eqnarray}
where $\tilde{r}\equiv\sqrt{x^{2}+y^{2}+z^{2}}$, as before. The first three terms correspond to the Schwarzschild geometry in the first order approximation. In fact, these expressions are a particular case of the general line-element for non-relativistic stationary sources, understood as a linear perturbation of Minkowski background spacetime \citep{Efstathiou}
\begin{eqnarray}
ds^{2}=c^{2}\left(1-\dfrac{2\Phi^{(g)}}{c^{2}}\right)dt^{2}-\left(1+\dfrac{2\Phi^{(g)}}{c^{2}}\right) \times
    \nonumber  \\
\times \left[ (dx^{1})^2+(dx^{2})^2+(dx^{3})^2\right]+2A^{(g)}_{i}dx^{i}dt,
\end{eqnarray}
with $g_{\alpha\beta}=\eta_{\alpha\beta}+h_{\alpha\beta}$,
where $\Phi_{(g)}$ and $A^{i}_{(g)}$ are the gravitoelectric and gravitomagnetic potentials respectively, which can be defined through the expressions below \citep{Efstathiou}
\begin{equation}
\bar{h}^{00}\equiv\dfrac{4\Phi_{(g)}}{c^{2}},\qquad \bar{h}^{0i}\equiv \dfrac{A^{i}_{(g)}}{c},
\end{equation}
and
\begin{eqnarray}
 A_{i}dx^{i}=\eta_{ij}A^{i}dx^{j}=-\vec{A}\cdot \vec{dx}, 
     \nonumber  \\ 
  \bar{h}_{\mu\nu}\equiv h_{\mu\nu}-\dfrac{1}{2}h_{\mu\nu}h,\qquad h=\eta^{\mu\nu}h_{\mu\nu}.
\end{eqnarray}
For a brief review on gravitolelectromagnetism in the perturbative as well as geometric approaches, see \cite{GravitomagnetismMashhoon:2003ax}. In this perspective, we can clearly see that the influence of the object's angular momentum in the surrounding spacetime express the gravitomagnetic effect (measured around Earth in the Gravity Probe B experiment \cite{Everitt:2011hp}).  

The PPN (Parameterized Post-Newtonian) formalism allows one to parameterize geometrical gravitational theories within the Solar System, or more generally around spherically symmetric stationary (possibly rotating) gravitational sources. Many dimensionless parameters appearing in Taylor expansions of the metric are thus defined describing deviations from GR. For example, two of the most important ones, $\beta$ and $\gamma$, can arise from an expansion in powers of $GM/c^{2}r$ of the most general static spherically symmetric spacetime \citep{Hartle}
\begin{eqnarray}
ds^{2}&=&D(r)c^{2}dt^{2}-B(r)dr^{2}-r^{2}\left(d\theta^{2}+\sin\theta^{2}d\varphi^{2}\right)
   \nonumber \\
&=&{\rm weak\: field\: (Newtonian)+post-Newt.},
\end{eqnarray}
obtaining
\begin{eqnarray}
D(r)&=&1-\Psi(r)+(\beta-\gamma)\Psi(r)^{2}+...,
    \nonumber  \\
 B(r)&=& 1+\gamma\Psi(r)+...,
\end{eqnarray}
where $\gamma$ is a measure of the spatial curvature produced by a unit rest mass, while $\beta$ is a measure of how much non-linearity is present in the superposition law for gravity. The higher order terms in the expansion give the so-called Post-Newtonian corrections while the expansion up to first order represent the weak field limit and is sometimes referred to as the Newtonian limit. However, strictly speaking there can be corrections to Newton gravity even in the first order (weak field) limit, as in the case of slow rotation where the gravitomagnetic potential and corresponding vector field allows non-Newtonian predictions, such as frame dragging or Lens-Thirring effects \citep{Efstathiou}. In GR, we have $\beta=\gamma=1$. Many of the so called gravitational classical tests can be expressed in terms of these parameters \citep{Efstathiou,Hartle}. For example, the deflection of light due to the Sun's gravitational field, the precession of the perihelion of planetary orbits and the Shapiro time delay of light signals in the Sun's field (see \cite{Hartle}, for example). Will \citep{Will:1971zza,Will:1972zz}, Ni \citep{Thorne:1970ws,Ni:1971wq} and Misner et al \cite{Misner:1974qy}
used ten parameters in the Beta-Delta notation and later, a different set of 10 parameters was used in the Alpha-Zeta notation (see for example \citep{Will:1981cz,Will:2005va}), but $\gamma$ and $\beta$ coincide in both notations. In the first notation $\Delta_{1}$ and $\Delta_{2}$ are intimately related to the off-diagonal elements characteristic of a Kerr-like metric, since $\Delta_{1}$ measures how much dragging of inertial frames is produced by unit linear momentum and $\Delta_{2}$ measures the difference between radial and transverse momentum on dragging of inertial frames.

All parameters are potentially useful to constrain alternative or extended geometric theories of gravity.  
For the slowly rotating object of our interest, when the expansions on the dimensionless quantities $GM/c^{2}r$ and $GJ/c^{3}r^{2}$ are taken up to first order (in the most general axially symmetric metric), one arrives at the following expressions
in spherical and (isotropic) Cartesian coordinates
\begin{widetext}
\begin{equation}
ds^{2}=c^{2}\left(1-\dfrac{2GM}{c^{2}r}\right)dt^{2}-\left(1+\gamma\dfrac{2GM}{c^{2}r}\right)dr^{2}-r^{2}\left(d\theta^{2}+\sin^{2}\theta d\varphi^{2}\right)+\left(1+\gamma+\dfrac{\alpha_{1}}{4} \right)\dfrac{2GJ}{c^{2}r}\sin ^{2}\theta d\varphi dt,\label{metricPPN}
\end{equation}
\begin{equation}
ds^{2}=c^{2}\left(1-\dfrac{2GM}{c^{2}\tilde{r}}\right)dt^{2}-\left(1+\gamma\dfrac{2GM}{c^{2}\tilde{r}}\right)\left( dx^{2}+dy^{2}+dz^{2}\right)+\left(1+\gamma+\dfrac{\alpha_{1}}{4}\right)\dfrac{2GJ}{c^{2}\tilde{r}^{3}}dt(xdy-ydx),\label{metricPPN2}
\end{equation}
\end{widetext}
respectively, where $\alpha_{1}\equiv 7\Delta_{1}+\Delta_{2}-4\gamma-4$, measures the extent of preferred frame effects and is equal to zero in GR. The following bounds taken from Will \cite{Will:2005va} due to local (Solar System) tests should be taken under consideration by any gravitational theory 
\begin{equation}
\vert \gamma-1\vert=2.3\times 10^{-5},\quad \vert \beta-1\vert=3\times 10^{-3},\quad \alpha_{1}< 10^{-4},
\nonumber
\end{equation}
where the first result was obtained through light deflection and time delay, the second is due to perihelion shift and the last from orbit polarization in Lunar Laser Ranging (LLR).

\subsubsection{Gravitomagnetic coupling between electric and magnetic fields}

We now consider Maxwell's inhomogeneous equations in the background spacetime around a spherical mass with slow rotation in the weak field (first order) limit (using Eq (\ref{metricPPN})). New gravitomagnetic terms appear in the Gauss and Maxwell-Amp\`{e}re laws due to the off-diagonal time-space metric component. The generalized Gauss law, Eq. (\ref{GaussGenerall}), has now a mixture of electric and magnetic components. 
The coupling to spacetime geometry and the gravitomagnetic effect provides magnetic corrections, which vanish in the non-rotating regime. This puts forward the interesting possibility of having even static magnetic fields as sources of electric fields.

A possible application concerns the magnetic field around rotating neutron stars. This field feels the presence of very strong gravity where the curvature of spacetime should not be neglected. In such astrophysical conditions the theory here exposed suggests an induced electric field component due to the coupling between the magnetic field and the gravitomagnetic character of gravity.

Another application comes from supermassive rotating black holes in the center of disk-shaped galaxies. It is well know that the frame-dragging character of the gravitomagnetic field around the rotating mass can be understood as a differential rotation of the curved spacetime around the rotation axis of the object (as seen from an observer far away from the source). Such a rotation is analogous to what happens in tornadoes where, in contrast to rigid bodies, the angular velocity is higher towards the center and decays with radial distance. Accordingly, from the coupling between gravitomagnetism and electromagnetic fields within Maxwell's equations, it is natural to expect a frame-dragging effect on these fields (see also \cite{Bicak:2006hs, Karas:2012mp}). Therefore, an electric field around a super-massive rotating black hole would feel the differential rotation of spacetime resulting in a spiral pattern for the electric field lines along the galactic equator. If this relativistic (non-Newtonian) effect might provide some light into the understanding of the formation processes of spiral structures in galaxies, remains up to this moment a challenging and open question.   

Changing the Cartesian (isotropic) coordinates in Eq. (\ref{metricPPN2}) to (axi-symmetric) cylindrical coordinates $(t,\tilde{R},\phi,z)$, the metric becomes
\begin{eqnarray}
ds^{2}&=&c^{2}\left(1-\dfrac{2GM}{c^{2}\left( \tilde{R}^{2}+z^{2}\right)^{1/2}}\right)dt^{2}
   \nonumber \\
&&-\left(1+\gamma\dfrac{2GM}{c^{2}\left( \tilde{R}^{2}+z^{2}\right)^{1/2}}\right)\left( d\tilde{R}^{2}+\tilde{R}^{2}d\phi^{2}+dz^{2}\right)
  \nonumber \\
&&+\left(1+\gamma+\dfrac{\alpha_{1}}{4}\right)\dfrac{2GJ}{c^{2}\left( \tilde{R}^{2}+z^{2}\right)^{3/2} }\tilde{R}^{2}dtd\phi,\label{metricPPN3}
\end{eqnarray}
where $\tilde{R}$ is a radial coordinate related to the physical distance to the rotation axis.
The Gauss law, in vacuum, for this case is
\begin{eqnarray}
\left[ \dfrac{1}{\tilde{R}}\partial_{\tilde{R}}(\tilde{R}D^{\tilde{R}})+\partial_{z}D^{z}+\dfrac{1}{\tilde{R}}\partial_{\phi}D^{\phi}\right]=0.
\end{eqnarray}
We will assume for simplicity no magnetic fields. Then we have the following linear (homogeneous and isotropic) constitutive relations \cite{FCFL}
\begin{equation}
D^{k}=\varepsilon_{0}\sqrt{-g}(g^{0i}g^{0k}-g^{00}g^{ik})E_{i} \,.\label{constituteppnaxial}
\end{equation}
Let us consider an electric field with axial symmetry [$\bold{E}=E_{\tilde{R}}(\tilde{R},z)\bold{e}^{\tilde{R}}+E_{\phi}(\tilde{R},z)\bold{e}^{\phi}$], where we have assumed that the $E_{z}$ component is negligible near the equatorial plane ($z=0$). Therefore, for this approximation, we have [$\bold{D}=D^{\tilde{R}}(\tilde{R},z)\bold{e}_{\tilde{R}}+D^{\phi}(\tilde{R},z)\bold{e}_{\phi}$]. Thus, the Gauss law provides the solution
$D^{\tilde{R}}(\tilde{R},z)=f(z)/\tilde{R}$.
 

\begin{widetext}
In the equatorial plane, setting $z=0$, the radial component of the electric field in the spacetime around the rotating massive object is given by 
%
%
\begin{equation}
E_{\tilde{R}}=\dfrac{f(0)}{4\varepsilon_{0}mc\tilde{R}^{2}}\dfrac{\left[r_{Sch}^{2}J^{2}(4+\alpha+4\gamma)^{2}+16c^{2}m^{2}(1-r_{Sch}/\tilde{R})(1+\gamma r_{Sch}/\tilde{R})\tilde{R}^{2} \right] ^{1/2}}{(1+\gamma r_{Sch}/\tilde{R})},\label{electric fieldppnslowrot}
\end{equation}
\end{widetext}
and for $J=0$
\begin{equation}
E_{\tilde{R}}=\dfrac{C}{\tilde{R}}\sqrt{\dfrac{(1-r_{Sch}/\tilde{R})}{(1+\gamma r_{Sch}/\tilde{R})}},
\end{equation}
where $C$ is a constant. On the other hand, the Faraday law gives $\partial_{z}E_{\phi}=\partial_{\tilde{R}}E_{\phi}=0$, therefore $E_{\phi}$ is an arbitrary constant that can be set to zero without a significant loss of generality.
We see that the gravitomagnetic term affects the gravitationally induced decay of the radial component of the electric field. 
 
 These astrophysical scenarios intend to illustrate possible gravitomagnetic effects, affecting electromagnetism directly through the very nature of the field equations in curved spacetime. In fact, in the most general case it becomes clear that electrostatics and magnetostatics are no longer separated, but instead become coupled in the presence of gravitomagnetism. We assumed for simplicity a vanishing magnetic field, but as previously mentioned, in general, even a static magnetic field will contribute to the electric field via this gravitomagnetic coupling induced by astrophysical sources with rotation.  

\subparagraph{Case of non-homogeneous and anisotropic electric permittivity:} 
Following our hypothesis that the electromagnetic properties of vacuum should reflect the local spacetime isometries, the constitutive relation (\ref{constituteppnaxial}) in this case (neglecting magnetic fields for simplicity) is generalized into
\begin{eqnarray}
D^{\tilde{R}}&=&-\varepsilon_{0}^{\tilde{R}}(\tilde{R},z)\sqrt{-g}g^{00}g^{\tilde{R}\tilde{R}}E_{\tilde{R}},\\
D^{\phi}&=&\varepsilon_{0}^{\phi}(\tilde{R},z)\sqrt{-g}\left[ (g^{0\phi})^{2}-g^{00}g^{\phi\phi}\right] E_{\phi}.
\end{eqnarray}
The important physical idea here is that the electric permitivity should have axial symmetry and depend on the direction, i.e., $(\varepsilon_{0})^{ij}={\rm diag}[\varepsilon_{0}^{\tilde{R}}(\tilde{R},z),\varepsilon_{0}^{\phi}(\tilde{R},z),\varepsilon_{0}^{z}(\tilde{R},z)]$. 

Let us assume as an Ansatz, the following expression
\begin{eqnarray}
\varepsilon_{0}^{\tilde{R}}=\varepsilon_{0}\left[ 1+\bar{\gamma}\dfrac{2GM}{c^{2}\tilde{R}}+\bar{\Delta}\dfrac{2GJ\tilde{R}}{c^{3}\left(\tilde{R}^{2}+z^{2}\right)^{3/2} }\right], 
\label{permitivitykerrppn}
\end{eqnarray}
where $\bar{\gamma}$ and $\bar{\Delta}$ are dimensionless parameters. This corresponds to the linear approximation of a Taylor expansion in powers of the relevant dimensionless quantities related to mass and angular momentum. By considering for simplicity the equatorial plane, i.e $z=0$, then the result in Eq. (\ref{electric fieldppnslowrot}) is generalized by replacing $\varepsilon_{0}$ with $\varepsilon_{0}^{\tilde{R}}(\tilde{R},z=0)$. As a consequence, in this case the electric field has a radial dependence which includes both the contribution from spacetime curvature (gravity) and a variable electric permittivity of vacuum. This is analogous to what we had in the spherical symmetric case generalized for the axial symmetric geometries. By turning off the angular momentum, we recover the spherical case. Further research is required to understand how the effects of having non-homogeneous (and anisotropic) permittivity and permeability tensors can have an impact on physical observables. Electric and magnetic fields interact with astrophysical plasmas in black hole accretion disks and around neutron stars. These interactions depend on the coupling constants which are basically the electric and magnetic properties of the medium. Such interactions need to be carefully taken care of, for instance, using magnetohydrodynamical computations. On the other hand, by changing the permittivity and permeability tensors, the propagation properties of electromagnetic waves is affected. We briefly discuss these issues in the conclusions. However, testing these hypotheses via observations need further analysis using the wave equations and also taking into account environmental effects, although this is not considered in the present work.


\section{Summary and Discussion}\label{conclusion}

Building on previous work \cite{FCFL}, in this paper, we explored the physical applications of electrodynamics in the background of a (pseudo) Riemann spacetime manifold. The main electromagnetic effects induced by spacetime curvature addressed here include the following: gravitational contributions for the decay of electric and magnetic fields in spherically symmetric spacetime, magnetic contributions to the Gauss law due to the gravitomagnetic character of the spacetime around rotating objects and the effects of relaxing the assumptions of homogeneity and isotropy of the electromagnetic properties of vacuum (the electromagnetic oscillations induced by gravitational waves were presented in another work \cite{FCFLGW}).
In particular, the physical (possibly measurable) effects of spacetime geometry in 
electromagnetic fields, expressed in terms of the extended Gauss and 
Maxwell-Amp\`{e}re laws, helps the comparison with the usual results obtained from electromagnetism in Minkowski spacetime, making clearer the physical interpretations of such effects. In the following, we briefly summarize the topics explored in this work:

{\it In the spacetime around spherical sources}, the results confirm that gravity induces a (geometric) contribution to the decay of electric and magnetic fields along any radial direction. This can be seen in Eqs. (\ref{electricSch}), (\ref{electricsch2}), (\ref{magneticfieldschw1}) and (\ref{magneticfieldschw2}). In principle, even electric fields due to plane charged plates, which are uniform in the absence of gravity will manifest a spatial variability as is clear from Eq. (\ref{electricnonunif}). This effect could be tested in principle, under appropriate experimental conditions similar to those used in the GP-B experiment, namely with recourse to drag-free motion of satellites in polar or equatorial orbits around a spherical mass, housing the probe (in this case a capacitor) in vacuum under extreme low temperatures achieved by cryogenics.

According to the hypothesis we explore here, in the curved geometry around a massive object the electric permittivity of vacuum is not homogeneous, but has a radial dependence instead. For black holes, this effect influences electric fields more strongly near the horizon. The strength of the effect is maximum at the horizon where the magnitude of the electric field is $0.5$ times weaker than the case with a constant permittivity tensor. 
In principle, this effect could  be tested using electric fields inside capacitors. The expressions (\ref{nonhomogeneityeffect1}) and (\ref{nonhomogeneityeffect2}) we obtained can be generalized to include the PPN approach. Similar considerations apply to magnetic fields around massive spherical (non-rotating) objects.

The hypothesis for non-homogeneous permittivity and permeability (electromagnetic) properties of vaccum leads to the result that the speed of electromagnetic waves is not the same in every point around a massive object. Instead, it must have a radial dependence. Using the Ansatz considered in this work, Eqs. (\ref{permitivityschw}) and (\ref{permeabilityschw}), in the first order approximation (with respect to Taylor expansions in powers of $r_{Sch}/r$), we get
\begin{equation}
c(r)=\dfrac{c_{0}}{(1+r_{Sch}/r)},\qquad c_{0}\equiv\dfrac{1}{\varepsilon_{0}\mu_{0}}. \nonumber 
\end{equation}
As a consequence, local observers could still agree about the velocity of light and the local conformal structure of spacetime, i.e., the local light-cone, but the change in the (local) light cone structure from   
one point to another now has both the influence of the spacetime curvature and the fact that the permittivity and permeability tensors change. This prediction for a non-homogeneous (but isotropic) speed of light in vacuum, in the spherical gravitational fields around massive objects, should have observable consequences that need to be tested experimentally.

{\it In the axisymmetric spacetime around rotating sources},
electrostatics and magnetostatics are no longer separated, but instead become coupled due to the presence of off-diagonal time-space metric components. We considered the metric in Eq. (\ref{metricPPN}) which corresponds to the spacetime around a rotating spherical mass in the weak field and slow rotation regime. This metric has the off diagonal component proportional to the angular momentum of the source. Such metric components, correspond in linearised gravity to the components of the gravitomagnetic potential characteristic of frame dragging (Lens-Thirring) effects around axisymmetric astrophysical systems in rotation. The coupling to spacetime geometry, in particular to gravitomagnetism, induces magnetic corrections to the Gauss law, i.e., Eq. (\ref{GaussGenerall}). This coupling suggests that even static magnetic fields can act as sources of electric fields via the gravitomagnetic frame dragging character of spacetime around rotating objects. In fact, the magnetic field around rotating neutron stars feels the presence of very strong gravity and therefore spacetime curvature should not be neglected. In such astrophysical conditions, the theory here exposed, suggests an induced electric field component, generated by the coupling between the magnetic field and the geometrodynamical character of gravity. Some work has been done in the past related to these issues (see \cite{Bicak:2006hs,Karas:2012mp}) but much more can be investigated. This is an illustration of a gravitomagnetic effect affecting electromagnetism directly through the very nature of the field equations in curved spacetime. It opens our perspectives in the way we imagine the astrophysical environment of such compact objects and also other sources of strong astromagnetic fields
 
As mentioned in Section \ref{secIIIB}, another possible application comes from supermassive rotating black holes in the center of disk-shaped galaxies. It is believed that the interaction of these supermassive black holes with the surrounding galactic environment is an important ingredient in the formation and evolution of the whole Galaxy and in the formation and evolution of AGNs and stellar formation bursts. On the other hand, it is also well known that the frame-dragging character of the gravitomagnetic field around a rotating mass can be understood as a differential rotation of curved spacetime around the rotation axis of the object. Such a rotation is analogous to what happens in tornadoes where, in contrast to rigid bodies, the angular velocity is higher towards the center and decays with radial distance. Accordingly, from the coupling between gravitomagnetism and electromagnetic fields within Maxwell's equations, it is natural to expect a frame-dragging effect on these fields. Therefore, an electric field produced by a supermassive charged, rotating black hole, would feel the differential rotation of spacetime resulting in a spiral pattern for the electric field lines along the galactic equator. This could induce currents and stationary charge density spiral patterns on the surrounding ionized gas. If this relativistic (non-Newtonian) effect might provide some light into the understanding of the formation processes of spiral structures in galaxies, remains a challenging and open question.

In any case, we found the radial component of an axially symmetric electric field, solution to the Gauss law, in the geometry given by Eq. (\ref{metricPPN3}). This solution again confirms that gravitational fields can decrease the magnitude of electromagnetic fields in vacuum and the gravitomagnetic term (proportional to the source's angular momentum) also contributes to this effect. 
It is also pedagogical to illustrate the role of the coupling between gravity and electromagnetism for testing different theories of gravity. We also generalized the constitutive relations in this case, to introduce non-homogeneity and anisotropy in the permittivity tensor, corresponding to a spacetime with axial symmetry.

In this work we briefly explored the idea that the electromagnetic properties of physical vacuum should follow the spacetime isometries. This idea reinforces the debate and research about the deep relation between gravity and electromagnetic fields, spacetime and the nature of vacuum. In particular, the conformal (casual) structure of spacetime is fundamentally connected to electromagnetism and the constitutive relations. The electromagnetic properties of vacuum and the conformal part of the metric appear in the constitutive relations between the fields and excitations, therefore, in principle the electromagnetic properties of vacuum should not be considered to be independent from the spacetime isometries or more specifically to the local conformal structure. Following this reasoning one can even reduce the number of concepts in an attempt to unify different approaches to the same problem. In this sense one can speak about the electromagnetic properties of the ``spacetime medium''.

Following these ideas, the (local) conformal structure of spacetime becomes a more fundamental concept than the metric structure. This can suggest the change of paradigm in spacetime physics according to which spacetime is not absolute but rather it is the local causal structure that is invariant. Therefore all observers agree on the (local) light cone related to some spacetime point but spacetime distances between any two events can be different for different observers. This change of paradigm naturally implies theories of gravity in which the (local) gauge symmetries are not Poincar\'{e} symmetries but rather those of the 15 parametric conformal group which includes the Weyl group and the proper conformal transformations. Dilatations (which belong to the Weyl group, together with the Poincar\'{e} transformations) and proper conformal transformations change the spacetime line element and obervers related by proper conformal transformations cannot be inertial observers. This wider gauge group naturally invites non-riemann geometries, in particular, general affine geometries with curvature, torsion and non-metricity tensors. 

{\it Regarding electromagnetic waves in the presence of gravity}, one can show that extra terms appear in the generalized wave equations which deserves further research. Indeed, going beyond the geometrical optics limit, light deflection (null geodesics) and gravitational redshift are not the only effects arising from the coupling between light and gravity. More generally, all electromagnetic waves can experience gravitational effects on the amplitudes, frequencies and polarizations \cite{FCFLGW} (see also \cite{Marklund:1999sp}). Besides, electric and magnetic wave dynamics become coupled in general, even in the Lorenz gauge.  
The coupling between the dynamics of different components suggests polarization effects. Moreover, there are terms in the wave equations depending on the first derivatives of the electromagnetic fields and similar terms for the wave equations written in terms of the potentials (\ref{wavepotential}). These terms might be responsible for a gravitational contribution to the decay of the oscillations, but formally, these are also compatible with the existence of  longitudinal modes induced by spacetime curvature, since similar terms appear in wave equations for vector fields with longitudinal modes. The fundamental reason behind this is the fact that in the presence of gravity the electric field in vacuum is no longer divergent-free (in the sense that $\partial^{k}E_{k}\neq 0$).

One can easily show that in the spacetime around a rotating astrophysical object, the equation for the potential  includes the influence of the gravitomagnetic term and therefore, one expects that the electromagnetic field will experience a frame-dragging (Lens-Thirring) effect due to the gravitomagnetism. In fact, one can show that the gauge invariant wave equation for the coupled electric and magnetic fields also includes similar gravitomagnetic terms. These terms not only contribute to the decay of the wave amplitude, but will also provide a geometrically induced coupling between the dynamics of the various electric and magnetic components which will most probably affect the polarization. Gravitomagnetic effects on electromagnetic waves deserve further research with potential applications for relativistic astrophysics related to Pulsars, AGNs and for the study of the electromagnetic counterpart of gravitational wave sources. \\

\section*{Acknowledgments}
FC acknowledges financial support of the Funda\c{c}\~{a}o para a Ci\^{e}ncia e Tecnologia (FCT) through the grant PD/BD/128017/2016 and Programa de Doutoramento FCT, PhD::SPACE Doctoral Network for Space Sciences (PD/00040/2012). FSNL acknowledges financial support of an Investigador FCT Research contract, with reference IF/00859/2012, funded by FCT/MCTES (Portugal). This article is based upon work from COST Action CA15117, supported by COST (European Cooperation in Science and Technology).



\end{document}